\documentclass[aps,prb,superscriptaddress,showkeys]{revtex4-2}
\usepackage[utf8]{inputenc}
\usepackage{amsmath,amssymb,amsfonts}
\usepackage{multirow}
\usepackage{graphicx,float}
\usepackage[colorlinks=true,
citecolor=blue,
linkcolor=blue,
urlcolor=blue]{hyperref}
\usepackage{footmisc}

\begin{document}

\title{Strain induced effects on the electronic and phononic properties of 2H and 1T$^{\prime}$ monolayer MoS$_{2}$}
\author{Saumen Chaudhuri}
\author{{A. K. Das}}
\affiliation{Department of Physics, Indian Institute of Technology Kharagpur, Kharagpur, 721302, India}
\author{{G. P. Das}}
\affiliation{Department of Metallurgical \& Materials Engineering, Indian Institute of Technology, Kharagpur, 721302, India}
\affiliation{Department of Physics, St. Xavier's College, 30 Mother Teresa Sarani, Kolkata 700016, India}
\author{B. N. Dev}
\email[corresponding author: ]{bhupen.dev@gmail.com}
\affiliation{Department of Physics and School of Nano Science and Technology, Indian Institute of Technology Kharagpur, Kharagpur, 721302, India}
\affiliation{Centre for Quantum Engineering, Research and Education, TCG Centres for Research and Education in Science and Technology, Sector V, Salt Lake, Kolkata 700091, India}

\begin{abstract}
First-principles calculations, within the framework of density functional theory, have been performed on the well-studied 2H and the less explored 1T$^{\prime}$ phase of single-layer MoS$_{2}$. We have addressed the strain-induced tunability of the electronic and phononic properties of both phases, and compared their stability against the applied strain. By considering a large number of strain profiles for both tensile and compressive stress, we have found that the electronic properties of both 2H and 1T$^{\prime}$ phases are sensitive to the direction of the applied strain and can be tuned in a controlled way. For the 2H phase, in most cases, a direct to indirect band gap transition at lower strain and a semiconductor to metal transition at higher strain is observed. The applied strain destroys the semimetallic nature of the 1T$^{\prime}$ phase via the overlapping of the bulk states with the topologically protected edge states. Significant strain-induced changes in the phononic properties, in the frequency of the phonon branches, as well as in the nature of the dispersion curves, are observed. A systematic change in the frequency of the optical phonon modes at the zone centre is seen for both phases. With increasing strain, the out-of-plane acoustic mode (ZA) turns imaginary, indicating a possibility of phase transition or instability of the crystal structure. The 2H phase appears to withstand a larger amount of strain and therefore possesses better stability compared to the 1T$^{\prime}$ phase since the imaginary branch starts to appear at much lower values of strain in the latter case. We highlight the significance of strain engineering in tuning the electronic and phononic properties and the safe limit of the strain application in different polymorphs of single-layer MoS$_{2}$.
\end{abstract}

\keywords{ First-principles study, strain, polymorphs, MoS$_{2}$, electronic and phononic properties}

\date{\today}
\maketitle

\section{Introduction}
Since the discovery of graphene \cite{novoselov2004electric} and its unique physical and chemical properties \cite{huang2011graphene, srivastava2012functionalized, yoo2011ultrathin, lin2009operation}, a lot of interest has been drawn towards atomically thin crystals \cite{novoselov2005two}. Within the continuously growing family of layered materials, transition metal dichalcogenides (TMDs) have attracted significant attention in the last decade due to their exceptional electronic and optical properties \cite{wilson1969transition, benameur2011visibility, radisavljevic2011single, pradhan2013intrinsic}, especially for their realization in the form of a single layer \cite{yazyev2015mos2} and their application in numerous fields of materials science \cite{radisavljevic2011integrated, wang2012electronics, johari2011tunable}. 

Within the large family of TMDs, a good amount of focus has been drawn towards MoS$_{2}$, due to its potential applicability in future nano-electronic devices \cite{yun2012thickness}. Similar to graphene or any other two-dimensional materials, single layers of MoS$_{2}$ can be produced from its bulk structure by use of mechanical or chemical exfoliation \cite{zeng2011single}, as well as by chemical vapour deposition (CVD), or physical vapour deposition (PVD) \cite{mohapatra2016strictly, wu2013vapor}. The crystal structure of MoS$_{2}$ consists of a Mo atomic plane sandwiched between two atomic planes of S atoms, with a strong intralayer covalent bonding and a much weaker interlayer van der Waals bonding \cite{wilson1969transition}, which makes its realization in the form of a single layer possible. Unlike graphene, due to the presence of three atomic layers, a monolayer of MoS$_{2}$ can exist in several polymorphs with different coordination of the Mo atom \cite{calandra2013chemically}, such as 2H, 1T and 1T$^{\prime}$. Due to the different structural forms, the polymorphs offer vastly different electronic and mechanical properties \cite{pizzochero2017point}. 2H-MoS$_{2}$, which is believed to be the most stable phase, is semiconducting in nature with a large direct band gap and ordered in ABA stacking sequence \cite{pizzochero2017point}. Whereas the 1T phase with ABC stacking order shows metallicity \cite{pizzochero2017point}. However, the 1T phase is unstable in pristine form and undergoes a dimerization distortion to result in 1T$^{\prime}$ phase \cite{calandra2013chemically, pizzochero2017point, hu2013new}, which consists of Mo-Mo zigzag chains along the y-direction. The 1T$^{\prime}$ phase has drawn much interest in recent times due to its exquisite electronic properties with a topologically non-trivial band structure, which arises due to the presence of band inversion \cite{pizzochero2017point, qian2014quantum}. The existence of strong spin-orbit coupling opens up a gap in the topologically protected edge states and thereby results in a quantum spin Hall (QSH) insulator phase \cite{qian2014quantum}.   

The tunability of the electronic, optical and mechanical properties of TMD nano-materials is highly desirable for their potential application in low-power flexible transistors, and numerous electromechanical and optomechanical systems \cite{chu2009strain}. The electronic properties of MoS$_{2}$ can be tuned by various approaches, such as varying the number of layers, inducing defects, applying an external electric field, or mechanical strain \cite{zhao2018study, komsa2015native, johari2012tuning}. Most of the approaches have limitations, as they irreversibly alter the material characteristics, thereby reducing the applicability of the material, whereas strain engineering can be an efficient route for its reversible application. Also, the application of strain is favoured over the other approaches due to its comparatively easy methodology. Strain can be induced by depositing the film over a stretchable substrate or by growing the film over a substrate with a large thermal expansion coefficient difference \cite{castellanos2012elastic, watson2021transfer, castellanos2013local}. So far, mechanical strain has been applied successfully in graphene \cite{huang2010probing} as well as other two-dimensional materials \cite{castellanos2012elastic, watson2021transfer, castellanos2013local} and the optical, phononic and electronic properties are tuned efficiently \cite{kumar2013mechanical}. 

Single-layer MoS$_{2}$ in 2H phase was found to be semiconducting with a large direct band gap \cite{pizzochero2017point, johari2012tuning}; tunability of this band gap and a possible semiconductor to metal transition with the application of strain might make monolayer 2H-MoS$_{2}$ suitable for application in tunable nanodevices. Similarly, for the 1T$^{\prime}$ phase, which exhibits QSH effect with a relatively large band gap \cite{qian2014quantum}, strain can be applied to tune the band gap or induce a topological phase transition \cite{lin2017topological}. Mechanical strains are almost invariably generated within the monolayer during its integration to any devices due to the different lattice structure, thermal expansion coefficient mismatch, and complex transfer processes \cite{castellanos2012elastic, watson2021transfer, castellanos2013local}. In order to understand the impact of such strain on the electronic properties and lattice dynamics, first-principles calculations are essential. There is a limit to which strain can be applied reliably before producing any damage to the structure. That limit can be predicted by calculating the phonon dispersion curves at different strain values. Although there have been several reports on strain-induced effects in MoS$_{2}$, they lack in phonon calculation; therefore, they did not predict the stability of the different crystalline phases or the maximum amount of strain that can be applied, which would not lead to dynamic instability of the structure \cite{johari2012tuning, kumar2013mechanical, lin2015insulator}. 

In this paper, we considered both tensile and compressive strains applied within the plane of 2H and 1T$^{\prime}$-MoS$_{2}$ along different directions and studied their impact on the electronic properties and lattice dynamics. To the best of our knowledge, no earlier reports \cite{kumar2013mechanical, lin2015insulator} considered such a wide variety of strain profiles and their impact on both the electronic and phononic properties of different polymorphs of monolayer MoS$_{2}$. We analyzed the strain-induced modifications in the electronic structures of both phases and correlated these modifications with the corresponding changes in the atomic structures. We then discussed the effect of strain on the phonon dispersion curves of the different crystalline phases. We predicted a stability range for all the strain profiles, thereby providing an idea about the strength of the different phases of single layer MoS$_{2}$ along certain directions. To correlate any modulations in the electronic and optical properties of single-layer MoS$_{2}$, found experimentally, with any possible local or global strain generated within the structure, these theoretical results would become very useful. 

\begin{figure}[h!]
 \centering
 \includegraphics[width=0.9\textwidth]{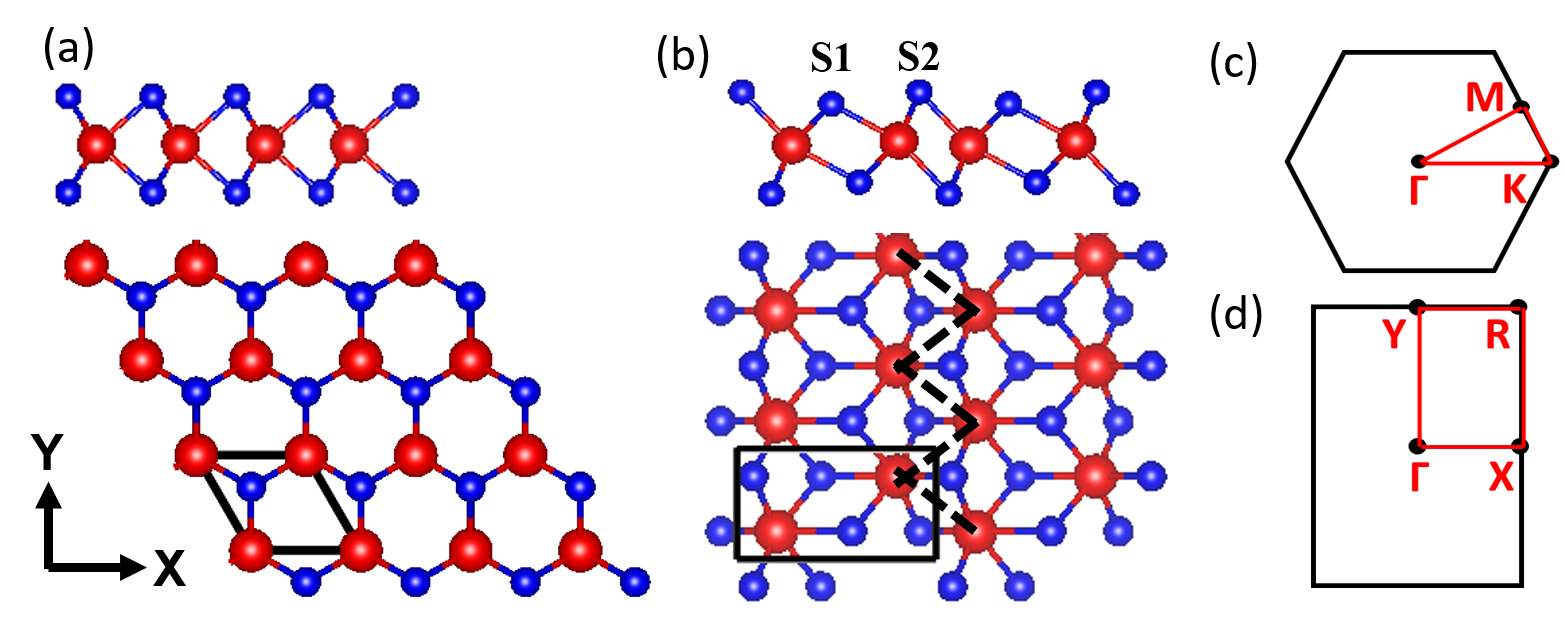}
\caption{Side (Upper panel) and top (Lower panel) views of the crystal structures of single-layer (a) 2H, (b) 1T$^{\prime}$-MoS$_{2}$. Mo and S atoms are in red and blue colour, respectively. Black solid lines show the unit cell and the black dashed lines in (b) represent the zigzag chain of Mo atoms. (c) and (d) show the Brillouin Zone (BZ) of 2H and 1T$^{\prime}$ MoS$_{2}$, respectively. The corresponding high symmetry points and paths are denoted in red. }
\label{Fig.1} 
\end{figure}

\section{Computational details}
In this work, first-principles calculations have been performed using ab-initio density functional theory (DFT) as implemented in the Vienna Ab Initio Simulation Package (VASP) \cite{kresse1996efficient, kresse1996efficiency} together with projector augmented wave potentials (PAW) to account for the electron-ion interactions \cite{kresse1999ultrasoft}. For the electronic exchange and correlation, generalized gradient approximation (GGA) with the Perdew–Burke–Ernzerhof (PBE) parametrization is adopted \cite{perdew1996generalized}. To simulate the layered structure, a supercell containing a vacuum of 20 {\AA} was used, which minimizes the interaction between the periodic images of the layers. After the application of strain, while the lattice parameters were kept fixed at the strained values, the internal positions of the atoms were optimized using a conjugate gradient algorithm until the forces on the atoms were less than 0.01 eV/{\AA} in each case. A well converged Monkhorst-Pack \cite{monkhorst1976special} k-points set of 21 $\times$ 21 $\times$ 1 together with a plane wave cutoff energy of 450 eV was used for the geometry relaxation and a much denser k-mesh of 41 $\times$ 41 $\times$ 1 was used for the post-relaxation calculations of the 2H phase. As the area of the BZ for the 1T$^{\prime}$ structure is half that of the 2H structure, the k-mesh was correspondingly scaled for the 1T$^{\prime}$ phase to keep the k-point density identical for both the phases. For the electronic structure calculation of 1T$^{\prime}$-MoS$_{2}$, spin-orbit coupling (SOC) is taken into account.  For the calculation of the phononic properties of single-layer MoS$_{2}$, density functional perturbation theory (DFPT) \cite{gonze1997dynamical} was employed, as implemented within VASP. Another tool, Phonopy \cite{togo2008first}, was used in order to extract the phonon frequencies from the DFPT results. To calculate the force constants and hence resulting phonon frequencies accurately, a large supercell of dimension 6 $\times$ 6 $\times$ 1 for the 2H- and 3 $\times$ 6 $\times$ 1 for the 1T$^{\prime}$-phase together with a strict energy convergence criterion of 10$^{-8}$ eV is used. The dimension of the supercells used for the phonon calculations is chosen carefully to produce well-converged results even in extreme strained cases. 
\begin{figure}[h!]
 \centering
 \includegraphics[width=0.9\textwidth]{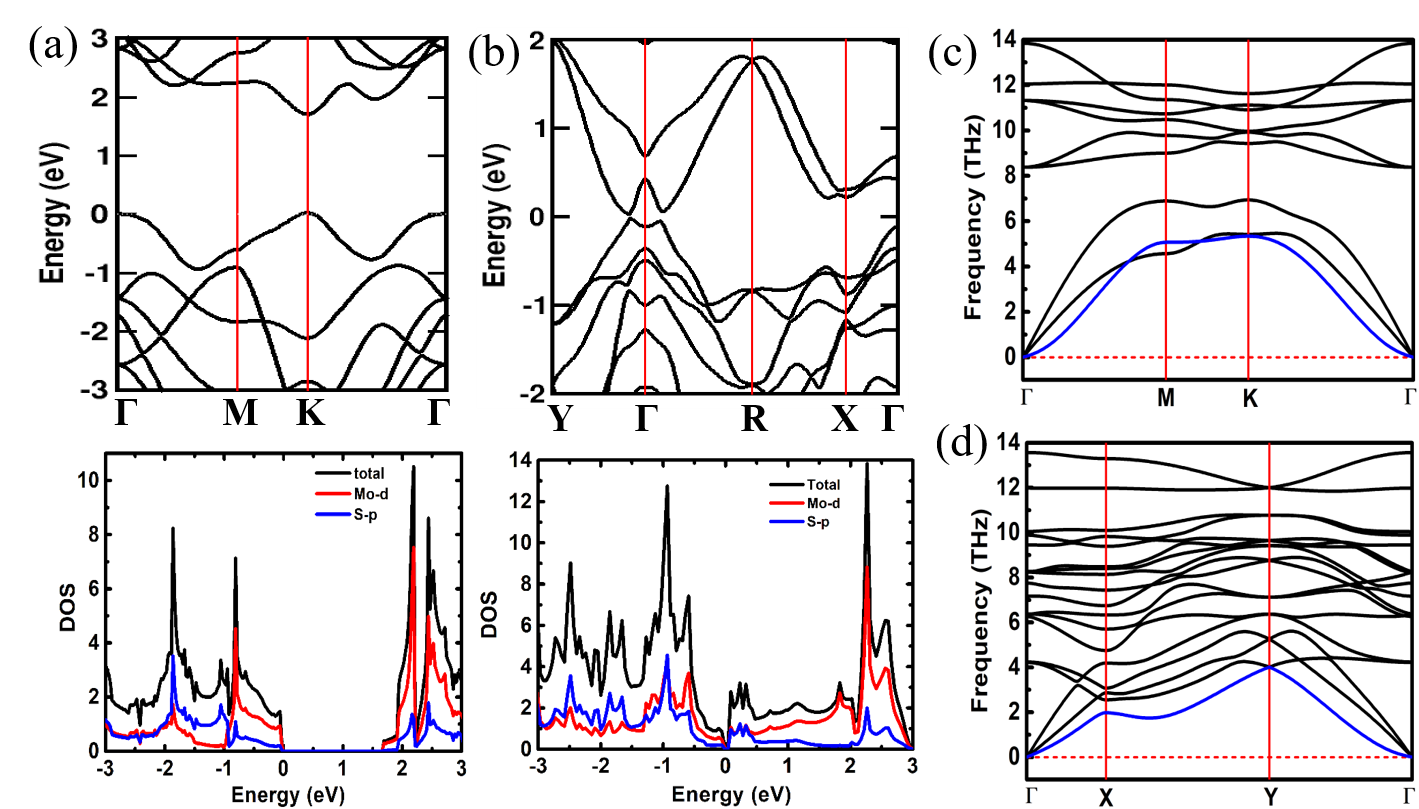}
 \caption{Electronic band structure (upper panel) and density of states (DOS) plots (lower panel) of (a) 2H and (b) 1T$^{\prime}$ monolayer MoS$_{2}$ in pristine condition. Fermi level is set at the top of the valence band. SOC is considered in the electronic structure calculation of 1T$^{\prime}$-MoS$_{2}$. Phonon dispersion curves of (c) 2H and (d) 1T$^{\prime}$ single layer MoS$_{2}$ in the unstrained condition. The out of plane acoustic (ZA) mode in the dispersion spectrum is coloured in blue.}
 \label{Fig.2}
\end{figure}

\section{Results and Discussion}
\subsection{Pristine MoS$_{2}$}
Before going into the details of strain-induced effects on single-layer MoS$_{2}$, we overview the calculated electronic and phononic properties of pristine monolayer MoS$_{2}$ in two different crystalline forms, 2H and 1T$^{\prime}$. Their crystal structures are shown in Fig.\ref{Fig.1}. In both the phases, 2H and 1T$^{\prime}$, the basic structure of monolayer MoS$_{2}$ remains the same, consisting of one atomic plane of Mo atoms sandwiched between two atomic planes of S atoms. However, the stacking order in these two phases is different. In the 2H phase, which is the most stable polymorph, the S and Mo atomic planes are arranged in ABA stacking sequence with the Mo atoms having trigonal prismatic coordination, as shown in Fig.\ref{Fig.1}(a). In the case of the 1T phase, which is the least stable phase, the S and Mo atomic planes are arranged in a rhombohedral ABC stacking sequence, with the Mo atoms having distorted octahedral coordination.  It is well documented that the 1T phase of monolayer MoS$_{2}$ is dynamically unstable in its pristine form with imaginary phonon branches in the dispersion spectrum throughout the Brillouin zone (BZ) \cite{calandra2013chemically, pizzochero2017point, hu2013new}. 1T phase single-layer MoS$_{2}$ undergoes a spontaneous dimerization distortion and results in 1T$^{\prime}$ MoS$_{2}$, where the Mo atoms form zigzag chains along the y-direction [Fig.\ref{Fig.1}(b)]. While in the 2H and the 1T phase monolayer MoS$_{2}$, there exists only one inequivalent S atom, the  1T$^{\prime}$ phase of MoS$_{2}$ contains two inequivalent S atoms with different Mo-S interatomic distances and hence inhomogeneous bond strengths. To distinguish these two S atoms, we label the S atom closer to (farther from) the Mo plane as S1 (S2), as shown in Fig.\ref{Fig.1}(b).

The electronic structure and the phonon dispersion curves of 2H and 1T$^{\prime}$ single-layer MoS$_{2}$ are shown in Fig.\ref{Fig.2}. The 2H phase is semiconducting with both the valence band maximum (VBM) and the conduction band minimum (CBM) appearing at the K point of the Brillouin Zone (BZ). This results in a direct band gap semiconductor with a band gap value of 1.68 eV, which agrees well with earlier reports \cite{johari2012tuning, kumar2013mechanical}. The band gap value is slightly underestimated compared to the experimentally found value of 1.9 eV \cite{lee2010anomalous, splendiani2010emerging}. Such an underestimation of the band gap is a well-known limitation within the DFT framework. From the DOS [Fig.\ref{Fig.2}(a)] and the orbital-projected band structure [Fig. S1], it is seen that both the conduction and the valence band edges are primarily composed of Mo-4\emph{d} and S-3\emph{p} atomic orbitals, albeit mainly Mo-4\emph{d} orbitals. In contrast to the semiconducting 2H-MoS$_{2}$, the 1T$^{\prime}$-MoS$_{2}$ monolayer has a semimetallic electronic structure with the VBM and the CBM located along the $\Gamma$ - Y line, as shown in Fig.\ref{Fig.2}(b). Due to the existence of band inversion at the $\Gamma$ -point, which can be seen from the orbital projected band structure shown in Fig.S2, and the strong spin-orbit coupling originated from the Mo atoms, there exists a band gap of 60 meV between the VBM and the CBM, which otherwise become degenerate if the spin-orbit coupling is not taken into account in the electronic structure calculation. 

Fig.\ref{Fig.2}(c) and (d) shows the phonon dispersion curves of single-layer 2H and 1T$^{\prime}$-MoS$_{2}$, respectively; they are in good agreement with previously reported curves \cite{qian2014quantum, li2012ideal}. The dispersion curves of 2H-MoS$_{2}$ consists of nine phonon branches, as the primitive unit cell contains three atoms, one Mo and two S atoms. In the dispersion curves, there exist three low energy acoustic branches, longitudinal acoustic (LA), transverse acoustic (TA) and one out of plane acoustic mode (ZA) or the flexural mode, which are well separated from the optical branches by a frequency difference of 1.38 THz or 46 cm$^{-1}$. While the LA and the TA modes show linear behaviour in the unstrained condition, the ZA mode shows quadratic behaviour near the zone centre. The 2H-MoS$_{2}$ monolayer has two Raman active modes, namely E$_{2g}$ and A$_{1g}$, at frequencies of 11.32 THz (377 cm$^{-1}$) and 12.04 THz (401 cm$^{-1}$), respectively. These frequency values agree well with the previously reported values \cite{mohapatra2016strictly, li2012ideal}. The atomic vibrations corresponding to the E$_{2g}$ and A$_{1g}$ modes are shown in Fig. S3. The phonon dispersion curves of 1T$^{\prime}$-MoS$_{2}$ are quite complicated with a large number of phonon branches, as can be seen in Fig.\ref{Fig.2}(d). The spectrum consists of eighteen phonon branches; three low energy acoustic and fifteen optical branches. Out of all the Raman active modes, our focus is particularly on four of them: J$_{1}$ (4.23 THz), J$_{2}$ (6.37 THz), J$_{3}$ (9.88 THz), and A$_{1g}$ (11.98 THz). Only these four modes show large enough intensities that corroborate with experiments \cite{pal2017chemically}.  The atomic vibrations corresponding to these four modes are shown in Fig. S3. 
\begin{figure}[h!]
 \centering
 \includegraphics[width=0.9\textwidth]{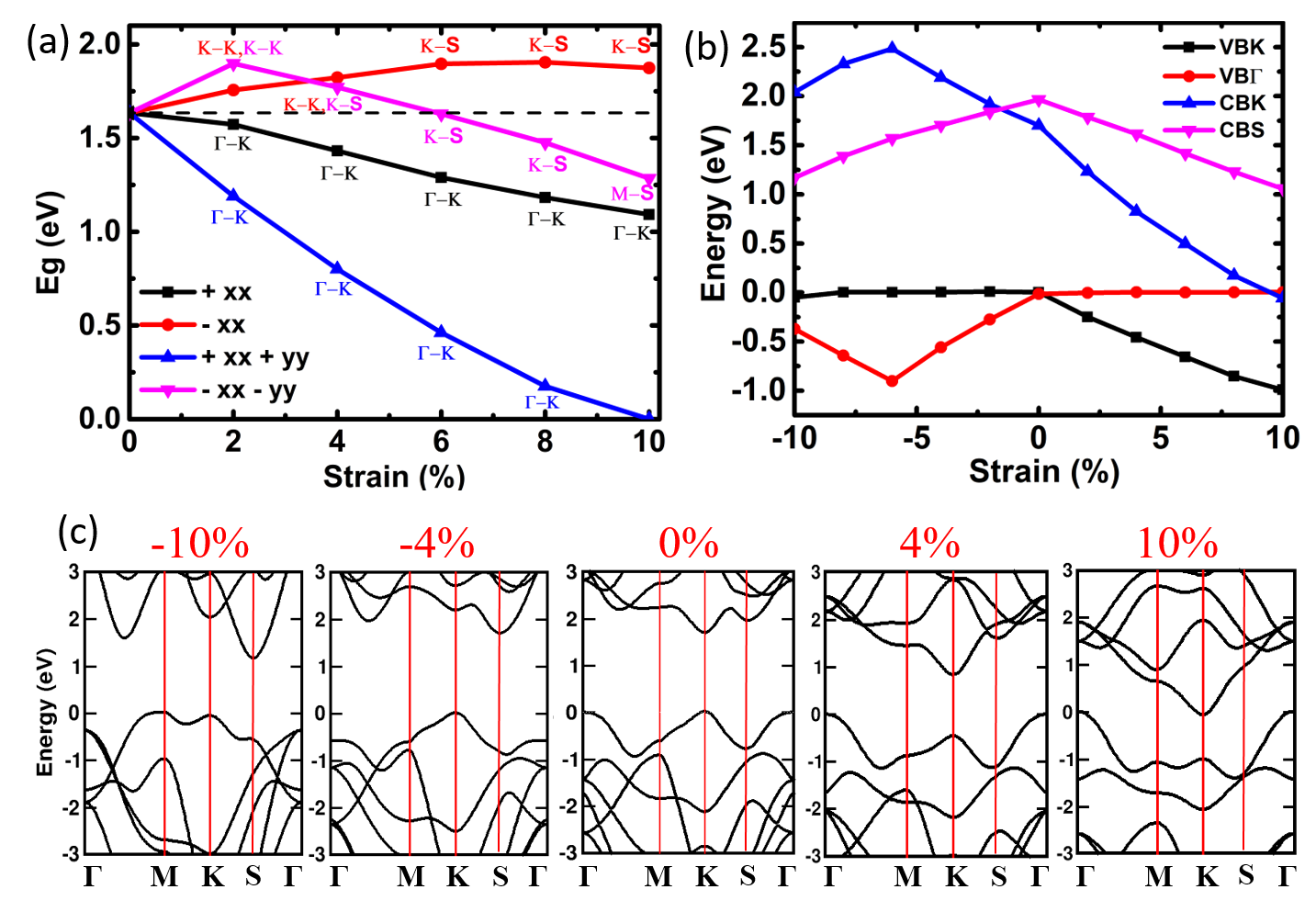}
 \caption{(a) Modulation of the band gap of 2H-MoS$_{2}$ under different strain profiles. The horizontal dashed line indicates the pristine band gap. The probable transitions are noted at each strain value with the corresponding colours. (b) and (c) are plots for the biaxial tensile and compressive strain. (b) Evolution of the most influential band edges in energy in reference to the Fermi level set at the top of the valence band and (c) Modification in the band structure at different values of strain. The strain values are noted at the top of the corresponding band structure plots.}
 \label{Fig.3}
\end{figure}

\subsection{Effect on the electronic properties}
To study the effect of mechanical strain on the electronic properties of monolayer MoS$_{2}$ in different crystalline phases, various in-plane strains, such as uniaxial tensile/compressive ($\pm$ XX, $\pm$ YY), biaxial tensile/compressive ($\pm$ XX $\pm$ YY) were applied on the relaxed geometry of the crystal structure, obtained from our calculations. The effect of strain on the band gap of 2H-MoS$_{2}$ is summarized in Fig.\ref{Fig.3}(a). For all the strain cases, the changes in the band gap have two features in common; a direct to indirect band gap transition and the ultimate lowering of the band gap, although the amount of change strictly depends on the direction of the strain application. The energy variation of the band edges, which are important in determining the valence band to conduction band transitions under biaxial tension and compression, is shown in Fig.\ref{Fig.3}(b). Note that, under the application of uniaxial (+XX) or biaxial (+XX+YY) tension, the magnitude of the band gap starts to decrease from very low values of strain. However, for uniaxial (-XX) or biaxial (-XX-YY) compression, the situation is different, i.e., the band gap value increases initially and then ultimately decreases. This is clear from Fig.\ref{Fig.3}(a) and (b). During compression, the CB edge at K shifts upward in energy while the VB edge remains fixed at K. As a result, the transition energy from the VB edge to the CB edge increases until the CB edge at K shifts to a point midway along the K-$\Gamma$ line at a strain value around 2\%. The point midway along the K-$\Gamma$ line is denoted as S-point in Fig.3(c). The changes in the band edge energies are monotonic except when the band edge at a specific k-point shift from one band to another. The systematic changes in the band structure of single-layer 2H- MoS$_{2}$ for some representative cases of biaxial strain are presented in Fig.\ref{Fig.3}(c), and those for the uniaxial strain are shown in Fig. S4. For both uniaxial and biaxial tension, the VBM is shifted from K to $\Gamma$ point at a strain around 2\%, and the CBM at K point downshifts in energy with increasing strain, keeping the band edges at other high symmetry points nearly unchanged. Thereby, the band gap decreases with increasing tensile strain. Although the modifications in the band structure are qualitatively the same for both the uniaxial and biaxial tension, the changes are more pronounced in the case of biaxial strain, with a rapid drop of CB edge at K point, which is clear from Fig.\ref{Fig.3}(b). For the biaxial tension, at a strain around 10\%, a semiconductor to metal transition with the closure of the band gap can be seen, as the CBM at K point starts to cross the Fermi level, which is clear from Fig.\ref{Fig.3}(b) and (c). In the case of compressive strain, the VBM remains at the K point, but the CBM gradually shifts from K to a point along the K-$\Gamma$ line. As the CB edge at the K point moves away from the Fermi level, an initial increase in the band gap is seen. With increasing compression, the VB edge at M point moves upward in energy for the biaxial case, and the CBM along the K-$\Gamma$ line shifts downward in energy for both the uniaxial and biaxial cases. Consequently, the band gap gradually decreases with increasing compression after an initial increase at low strain values. This is summarized in Fig.\ref{Fig.3}(b).
\begin{figure}[h!]
 \centering
 \includegraphics[width=0.9\textwidth]{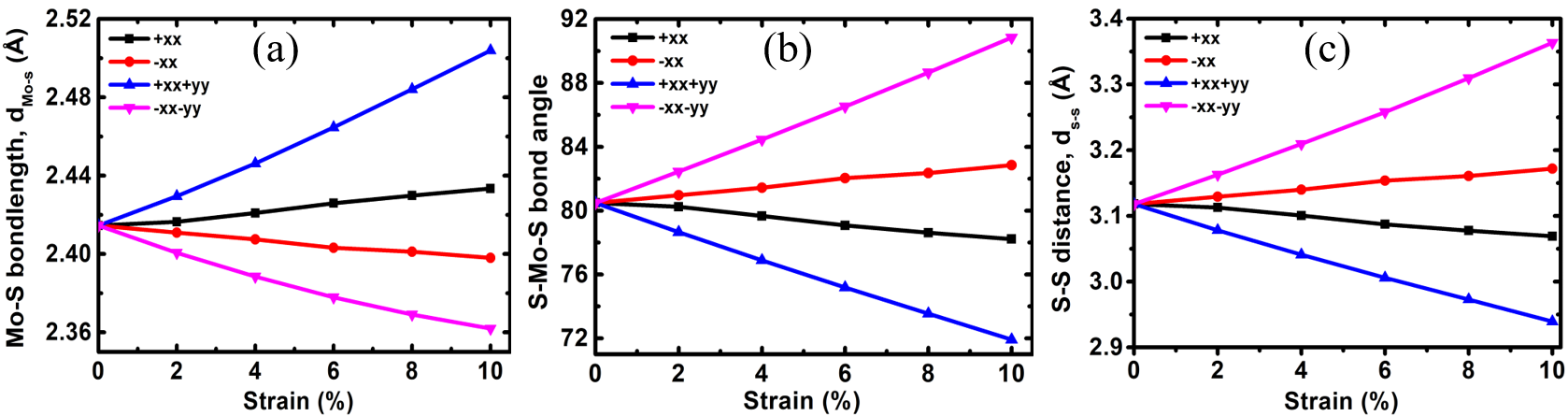}
 \caption{Strain-induced modifications in the (a) Mo-S bond length, (b) bond angle (in degree) between the Mo atom and the S atoms from two successive S planes and (c) S-S interplanar distance of 2H-MoS$_{2}$. }
 \label{Fig.4}
\end{figure}

Under the application of strain, the changes in the structural parameters of 2H-MoS$_{2}$, such as Mo-S bond length, S-Mo-S bond angle and the S-S interplanar distance, are shown in Fig.\ref{Fig.4}. The structural parameters calculated for 2H-MoS$_{2}$ in the unstrained condition are found to be in good agreement with earlier reports \cite{johari2012tuning}. As can be seen from Fig.\ref{Fig.4}, with the application of tensile (compressive) strain, the Mo-S bond length increases (decreases) with the decrease (increase) in both the S-Mo-S bond angle and the S-S interplanar distance. For both tension and compression, the changes in the structural parameters are found to be more pronounced for the biaxial cases than for the uniaxial ones. After the application of strain, the strained in-plane lattice parameters were kept fixed, so the atoms would naturally tend to relax in the out-of-plane direction due to Poisson contraction or expansion and, therefore, the S-Mo-S bond angle and the S-S interplanar distance change. The changes in the electronic structure with the application of strain can be easily understood in terms of the changes in the structural parameters and the resulting changes in the interaction strength of different atomic orbitals. As can be seen from Fig. S1, the VB edge at K point is mainly composed of d$_{xy}$/d$_{x^{2}-y^{2}}$ orbitals of Mo, whereas the VB edge at $\Gamma$ and the CB edge at K point are composed of Mo-d$_{z^{2}}$ and S-p$_{z}$ orbital in antibonding state and Mo- d$_{z^{2}}$ orbital, respectively. In the case of tensile strain, the Mo-S bond elongates, and the S-Mo-S bond angle reduces; thus, the in-plane interaction of d$_{xy}$/d$_{x^{2}-y^{2}}$ orbital strengthens, but the out-of-plane interaction between Mo- d$_{z^{2}}$ and S-p$_{z}$ orbital weakens. Thus the antibonding Mo- d$_{z^{2}}$ state at the CB edge at K falls in energy, and this explains the observed energy shifts in the band edges of 2H-MoS$_{2}$ single-layer with the application of strain.

\begin{figure}[h!]
 \centering
 \includegraphics[width=0.9\textwidth]{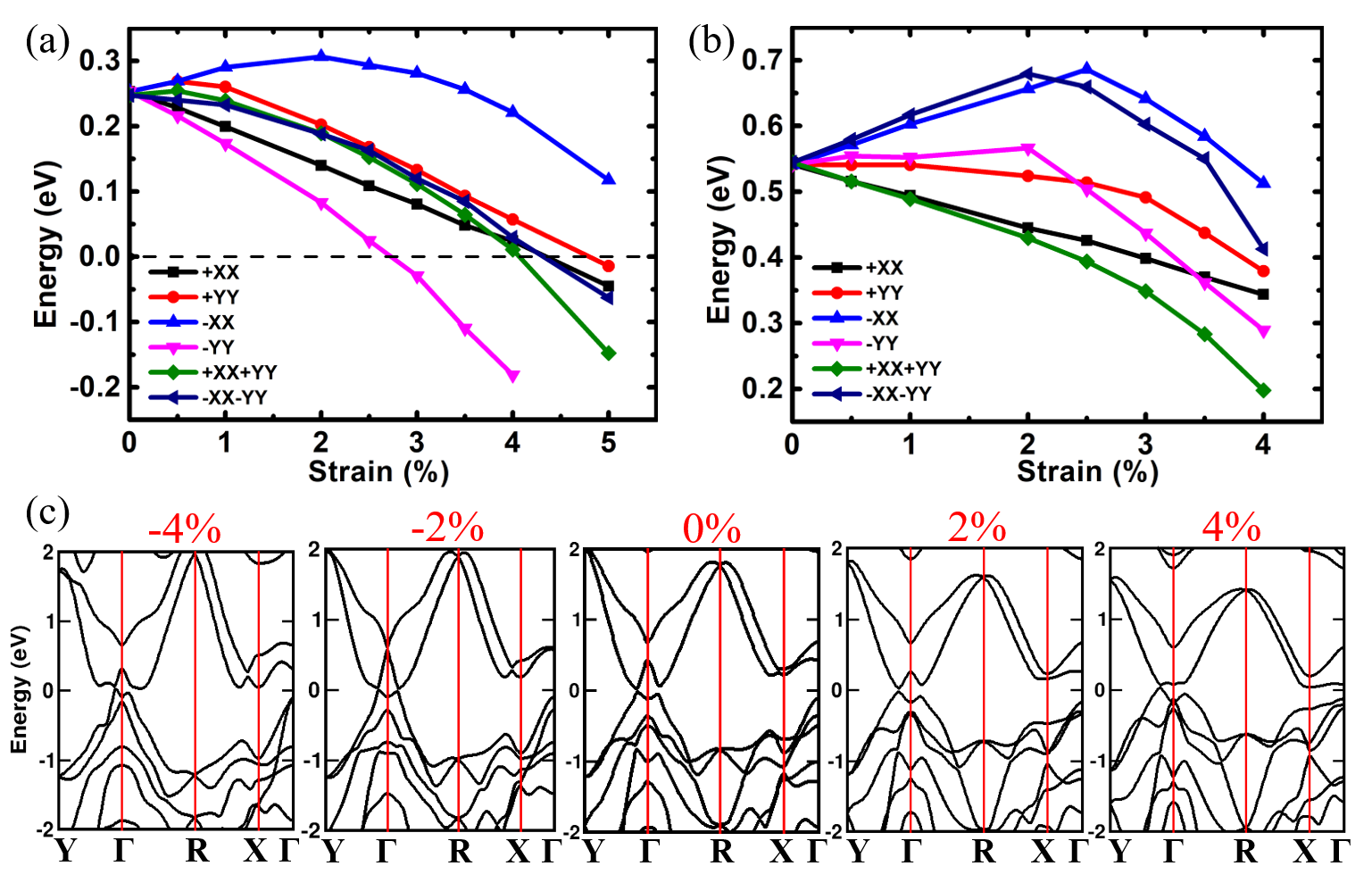}
 \caption{Strain-induced variation in (a) the band gap between the VB edge along Y-$\Gamma$ line and the CB edge at X, (b) the inverted band gap at $\Gamma$ of 1T$^{\prime}$-MoS$_{2}$. (c) Systematic changes in the band structure of 1T$^{\prime}$-MoS$_{2}$ at different values of biaxial tensile and compressive strain. The strain values are noted at the top of the corresponding band structure plots.}
 \label{Fig.5}
\end{figure}

The strain-induced changes in the electronic properties of single-layer 1T$^{\prime}$-MoS$_{2}$, which exhibit a non-trivial band topology in its unstrained condition, are summarized in Fig.\ref{Fig.5}. Here, we chose a smaller range of strain values for reasons explained later in section C. To understand the effect of strain in detail, we have applied both tensile and compressive in-plane strains in all possible directions (uniaxial and biaxial along X and Y). The evolution of two essential features of the band structure of 1T$^{\prime}$-MoS$_{2}$ is shown in Fig.\ref{Fig.5}(a) and (b). In Fig.\ref{Fig.5}(a), the band gap between the CB edge at X and the VB edge along the Y-$\Gamma$ line, originated from the edge states, is shown. The inverted band gap at $\Gamma$ point between the conduction band and the valence band originated from the bulk states is shown in Fig.\ref{Fig.5}(b). These changes in the band gap and the band edge states show how the bulk states might affect the conduction through the topologically protected edge states by overlapping with them and eventually with the Fermi level when the structure is strained. The band gap between the VB edge along the Y-$\Gamma$ line and the CB edge at X gradually decreases, and a transition from insulator to metal occurs for both tensile and compressive strain, as can be seen from Fig.\ref{Fig.5}(a). The inverted band gap at $\Gamma$-point can also be seen to decrease for both tension and compression with different trends, as shown in Fig.\ref{Fig.5}(b). For all the strain profiles calculated, depending upon the direction of strain application, the most notable changes in the band structure seem to happen at the CB edge at X and VB edge at $\Gamma$ point. The shape of the VBM and the CBM, originating from the edge states, remains nearly unchanged with a band gap value not less than 40 meV for a strain up to $\pm${3}\%. The strain-induced changes in the band structure for some representative values of biaxial strain are shown in Fig.\ref{Fig.5}(c), and those for the uniaxial strain are shown in Fig. S5 and S6. For uniaxial tension along x, the VB edge at $\Gamma$ point starts to move upward in energy and becomes the VBM at a strain of 4\%. The band structure calculations show that uniaxial tensile strain along x is more effective in terms of the insulator to metal transition (IMT), and strain along y becomes more effective for uniaxial compressive strain. These can also be seen from the band structure plots presented in Fig. S5 and S6 and from the band edge modulations shown in Fig.\ref{Fig.5}(a). The interesting point to note here is that, while the biaxial strain is the most effective strain profile in producing the IMT in the 2H phase, it becomes relatively less effective for the IMT in 1T$^{\prime}$ phase compared to the uniaxial cases. This less effectiveness is because the biaxial strain on 1T$^{\prime}$-MoS$_{2}$ not only changes the CB edges but also lowers the VB edges in energy. Therefore, the IMT occurs at higher biaxial strain values than the uniaxial cases. 

\begin{figure}[h!]
 \centering
 \includegraphics[width=0.9\textwidth]{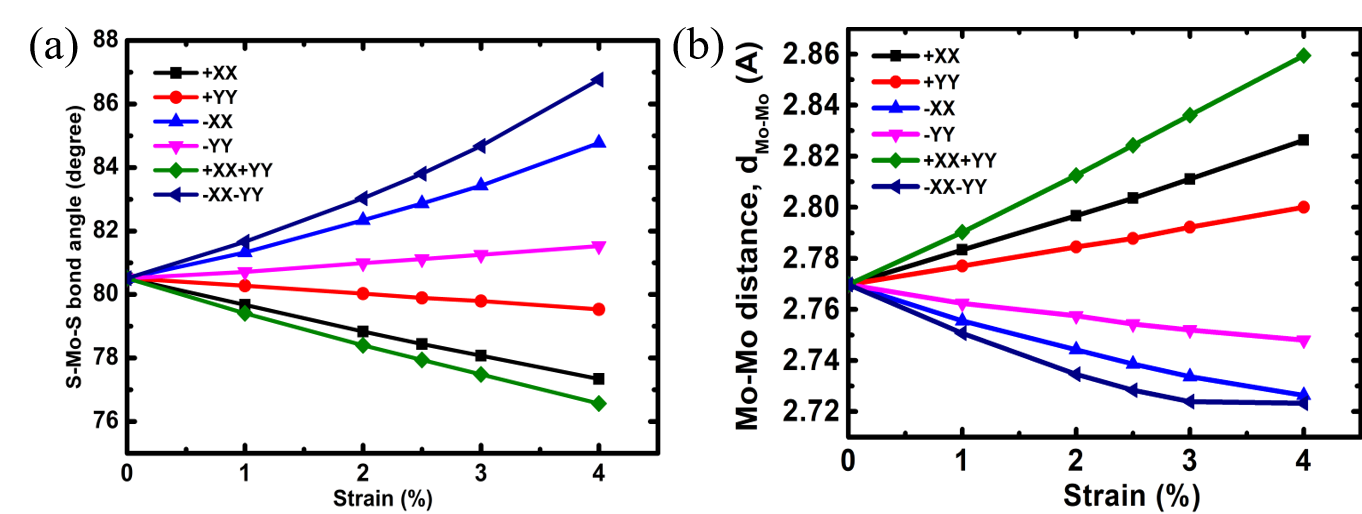}
 \caption{Strain-induced changes in the structural parameters of 1T$^{\prime}$-MoS$_{2}$. (a) The bond angle between the Mo atom and the S1 atoms from two successive S planes. (b) The Mo-Mo nearest neighbour distance along the zigzag chain, i.e. along y-direction.}
 \label{Fig.6}
\end{figure}

From the orbital projected band structure calculation of 1T$^{\prime}$-MoS$_{2}$, as shown in Fig. S2, it is seen that the CB edge at X point is mainly composed of Mo-d$_{x^{2}-y^{2}}$ orbitals with antibonding character and the VB edge at $\Gamma$ point is composed of the combination of S-p$_{y}$ and Mo-d$_{xy}$/d$_{xz}$/d$_{yz}$ orbitals. The insulator to metal transition with tensile or compressive strain is mainly driven by the energy lowering of the CB edge at X point. The strain-induced changes in the structural parameters of 1T$^{\prime}$-MoS$_{2}$, such as the Mo-Mo distance along the zigzag chain and the S1-Mo-S1 bond angle, are shown in Fig.\ref{Fig.6}. In the unstrained condition, the Mo-Mo bond distance along the zigzag direction is found to be 2.77 {\AA}   with a much larger projection of the bond along the x-axis than y, which agrees well with earlier reports \cite{lin2015insulator}. Due to the larger projection of the bond along the x-axis, naturally, the bond will be weaker and more sensitive to any perturbation along the x-axis than the y-axis. The application of strain will alter the Mo-Mo distance, and therefore, the in-plane interaction between Mo-d$_{x^{2}-y^{2}}$ orbitals will be changed. Thus, the CB edge at X, which is composed of Mo-d$_{x^{2}-y^{2}}$ orbitals, changes in energy following the changes in the Mo-Mo bond distance. Strain along the x-axis is expected to change the Mo-Mo bond length much effectively than strain along the y-axis due to the larger projection of the bond along x, which is also evident from Fig.\ref{Fig.6}(b). Similarly, one can also understand the change in energy of the VB edge at $\Gamma$ point from the change in bond angle between S1 and Mo atoms with strain and the resulting change in the interaction strength between the S-p$_{y}$ and the Mo-d$_{xy}$/d$_{xz}$/d$_{yz}$ orbitals. 

\begin{figure}[h!]
 \centering
 \includegraphics[width=0.9\textwidth]{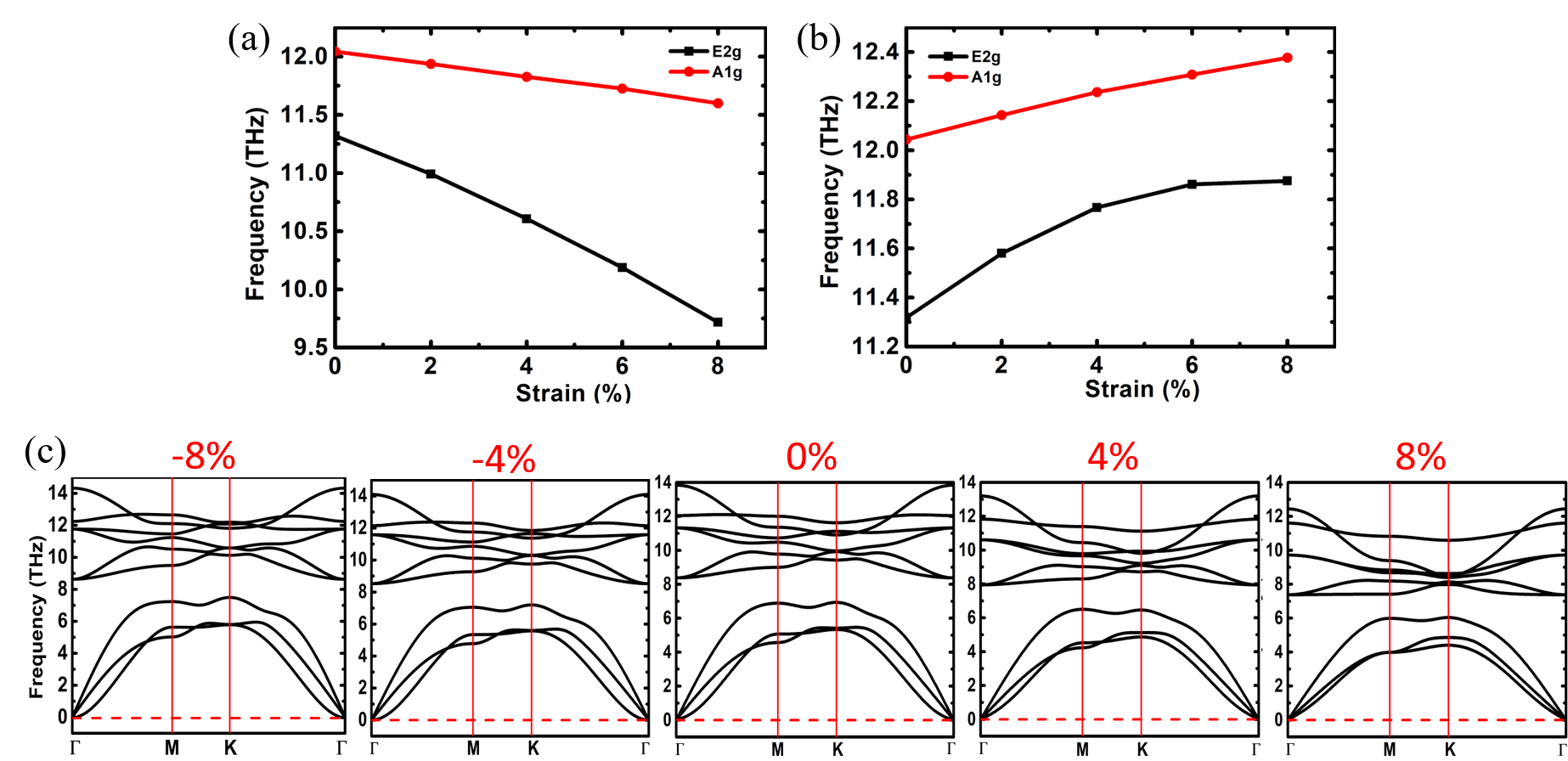}
 \caption{Modifications in the frequency of the Raman active E$_{2g}$ and A$_{1g}$ modes of 2H-MoS$_{2}$ under (a) biaxial tensile and (b) biaxial compressive strain. (c) Systematic changes in the phonon dispersion curves under biaxial tensile and compressive strain. The zero-frequency level is denoted in the spectrum by the red dashed line.}
 \label{Fig.7}
\end{figure}

\subsection{Effect on the phononic properties}
After assessing the strain-induced modifications in the electronic properties of single-layer 2H and 1T$^{\prime}$-MoS$_{2}$, we now turn our focus on how the application of strain may alter the lattice dynamics. To understand the dynamic behaviour of monolayer MoS$_{2}$ polymorphs and to find the critical strain before which the MoS$_{2}$ sheet becomes dynamically unstable, we applied strain biaxially on 2H-MoS$_{2}$ and both biaxially and uniaxially on 1T$^{\prime}$-MoS$_{2}$ single layer. The biaxial tensile- and compressive-strain-induced modifications in the frequency of the Raman active E$_{2g}$ and A$_{1g}$ modes of 2H phase are shown in Fig.\ref{Fig.7}(a) and (b), respectively. With increasing tensile (compressive) strain, the frequency of both the E$_{2g}$ and A$_{1g}$ modes decreases (increases) significantly at a rate of 6.67 cm$^{-1}$ (2.32 cm$^{-1}$) and 1.85 cm$^{-1}$ (1.38 cm$^{-1}$) per \% strain, respectively. Such softening or hardening of phonon modes with tensile or compressive strain can be understood following the microscopic definition of the phonon mode Gr\"{u}neisen parameter \cite{angel2019stress, conley2013bandgap}. This states that a positive strain, i.e. a tensile strain, would lead to the decrease of the frequency of a phonon mode for material with a positive mode Grüneisen parameter and would therefore result in the softening of the phonon mode, or, correspondingly, in the hardening of the phonon mode in case of compressive strain. Similar trends in the frequency of the phonon modes with applied strain have also been observed in graphene, and other two-dimensional materials, both in experiment and theory \cite{conley2013bandgap, rakshit2010absence, jha2014strain, mohiuddin2009uniaxial}.  The different rates of softening or hardening of these two modes with strain can be understood from the fact that an application of in-plane strain is more effective in stretching or compressing the in-plane bonds rather than the out-of-plane bonds. Hence the in-plane strain affects the dynamics of the in-plane phonon modes much strongly than the out-of-plane phonon modes. Some of the representative phonon dispersion spectra of 2H-MoS$_{2}$ at different values of biaxial tensile and compressive strain are shown in Fig.\ref{Fig.7}(c). With increasing tensile strain, the out-of-plane acoustic mode (ZA) or the flexural mode tends to become linear, which shows a quadratic behaviour in the unstrained condition. The frequency gap between the optical and acoustic branches seems to decrease with tensile and compressive strain. Absence of imaginary modes under tensile or compressive strains at least up to 8\%, as can be seen from Fig.\ref{Fig.7}(c), guarantees the stability of the structure within the applied strain range. Note here that the calculation of the phonon dispersion curves is done with a large supercell of dimension 6 $\times$ 6 $\times$ 1. Such a large supercell is required to consider the contribution from the long-range interaction between distant atoms towards the summation of the force constants and thereby ensure a well-converged result for the strained structures. The emergence of imaginary modes in the phonon dispersion curves is significant as they indicate instability of the crystal structure and thereby create a stability limit. Calculations beyond that strain limit are unreliable. Also, this sets a limit on applied strain in experiments. Although the modifications in the phonon dispersion curves under uniaxial strain are not shown here, we expect the structure to remain stable at least within the stability range of the biaxial strains since the modifications in the structural parameters under uniaxial strain are qualitatively same but quantitatively much smaller compared to the biaxial cases.  

\begin{figure}[h!]
 \centering
 \includegraphics[width=0.9\textwidth]{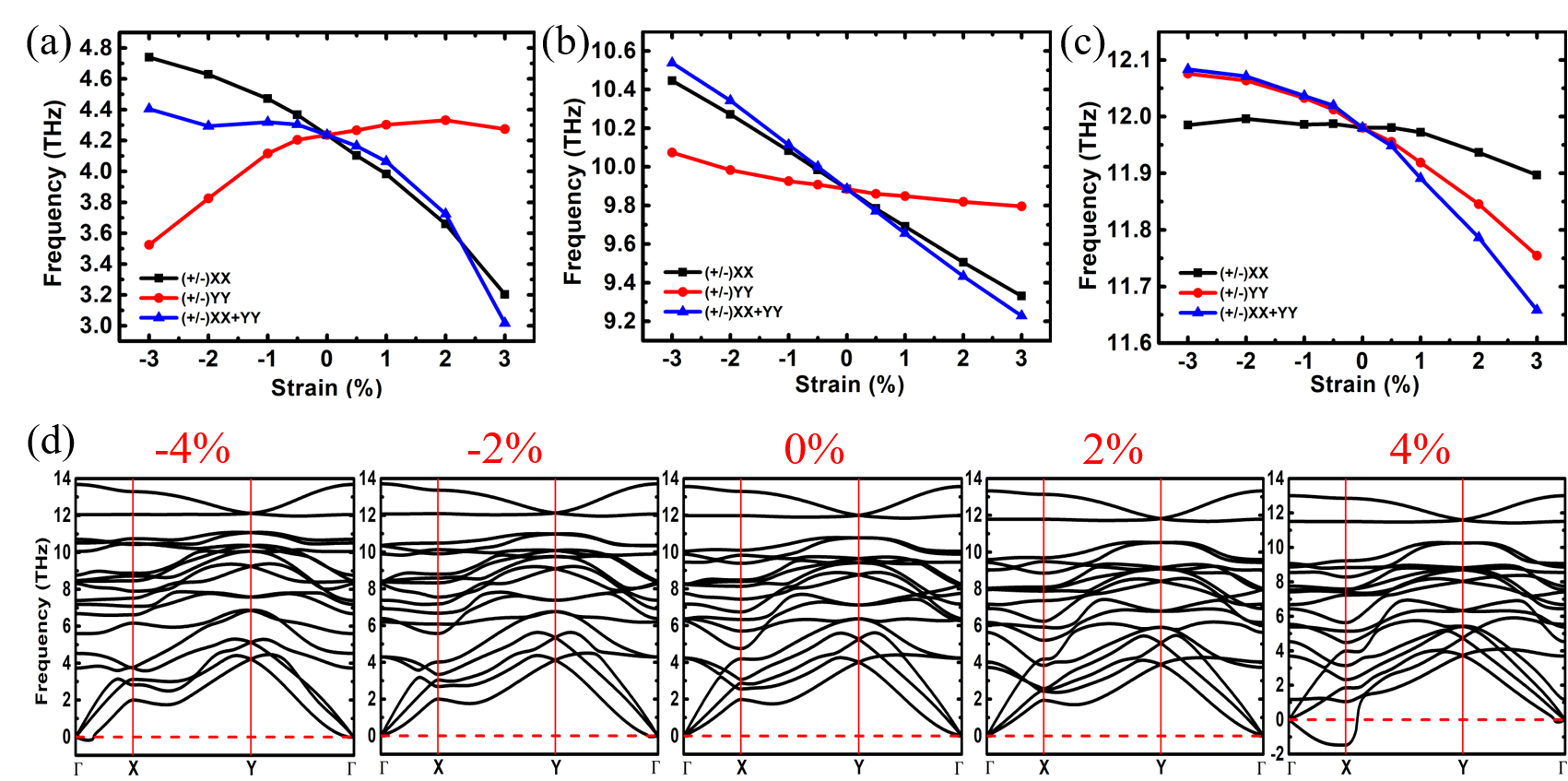}
 \caption{Modifications in the frequency of the Raman active (a) J$_{1}$, (b) J$_{3}$ and (c) A$_{1g}$ modes of 1T$^{\prime}$-MoS$_{2}$ under different strain profiles. (d) Systematic changes in the phonon dispersion curves under biaxial tensile and compressive strain. The zero-frequency level is denoted in the spectrum by the red dashed line.}
 \label{Fig.8}
\end{figure}

We then went on to investigate the strain-induced effects on the phonon dispersion curves of single layer 1T$^{\prime}$-MoS$_{2}$. The strain-mediated changes in the frequency of the Raman modes J$_{1}$, J$_{3}$ and A$_{1g}$ are shown in Fig.\ref{Fig.8}. Similar to the case of 2H-MoS$_{2}$, most of the phonon modes soften with the application of tensile strain and harden with compressive strain. It can be seen from Fig.\ref{Fig.8} that, in the case of uniaxial strain along the y-axis, the strain-induced changes in the Raman modes, specifically in J$_{1}$ and J$_{3}$ mode, are not only subtle but also different compared to the changes induced by uniaxial strain along the x-axis or by biaxial strain. Such direction-dependent changes in the frequency of the phonon modes can be understood by examining the eigenvector corresponding to the phonon mode. Consider the J$_{1}$ mode, for example, which has atomic vibrations along the y-axis only, with neighbouring atoms vibrating in opposite phases. Therefore, it is expected that strain along x or biaxial strain would strongly affect the vibrational symmetry and hence the vibrational frequency of the phonon mode, compared to the strain along the y-axis. This is clear from Fig.\ref{Fig.8}(a). Also, the tensile or compressive strain along the y-axis is the least effective among all the strain profiles considered in altering the geometrical parameters such as the S-Mo-S bond angle and the Mo-Mo nearest neighbour distance. This explains the lesser modification in the frequency of the J$_{1}$ and J$_{3}$ mode under the application of strain along the y-axis. The phonon dispersion curves of 1T$^{\prime}$-MoS$_{2}$ monolayer at some representative values of biaxial tensile and compressive strain are presented in Fig.\ref{Fig.8}(d). Note that, in the cases of uniaxial and biaxial strain, an imaginary mode emerges into the dispersion curves of 1T$^{\prime}$ structure at a much lower strain compared to the 2H cases. From Fig.\ref{Fig.8}(d), it can be seen that at 4\% biaxial tensile or compressive strain, the ZA mode turns imaginary. For the uniaxial strain along the y-direction, no imaginary branch is seen in the dispersion curves at the same amount of strain. This indicates a larger strength of the structure along y, which happens to be the direction of the zigzag chain of Mo atoms, compared to other directions. This comparatively larger strength along the y-direction is possibly because strain along the y-direction distorts the lattice structure rather slowly. From the variation of the S-Mo-S bond angle with tensile strain (see Fig.\ref{Fig.6}(a)), it can be seen that the bond angle decreases with increasing strain. The bond angle decreases the most with the biaxial strain, and the uniaxial strain along x nearly follows the same results. In contrast, the decrease in the bond angle with uniaxial strain along y is much smaller. From the phonon calculations, it is observed that, at 4\% strain, only the structure strained uniaxially along the y-direction remains stable with no imaginary phonon modes. For the other two cases, namely, the biaxial and the uniaxial strain along x, at 4\% strain, the out-of-plane acoustic mode or the ZA mode becomes imaginary, indicating instability of the structure. This suggests that there may exist a lower limit of the S-Mo-S bond angle above which the structure remains stable and for the 1T$^{\prime}$ structure, a bond angle value of $78 ^{\circ}$ can be taken as the lower limit. At 4\% strain, only the structure strained uniaxially along the y-direction possesses a bond angle that is greater than the lower limit of stability; this apparently produces a stable structure with no imaginary phonon modes. A clear understanding regarding the appearance of the imaginary modes during compression is lacking. However, an upper limit of the bond angle may exist, which may dictate the stability of the structure under compressive strain. There may be some other possibilities as well, such as the formation of ripple or warping of the monolayer structure. The softening of the ZA phonon mode under compressive strain (see Fig.\ref{Fig.8}(d)) may indicate the rippling of the monolayer sheet. For graphene, the hardening or softening of the ZA mode under strain application is generally associated with the non-occurrence or occurrence of rippling \cite{rakshit2010absence}. The emergence of imaginary branches in the phonon dispersion curves at certain strain values suggest a probable phase transition or dynamic instability of the structure. It is expected that single layer 1T$^{\prime}$-MoS$_{2}$ cannot withstand a large amount of strain due to its distorted dimerized crystal structure in its pristine condition. The appearance of the imaginary phonon modes in monolayer 1T$^{\prime}$-MoS$_{2}$ at much lower values of strain compared to the 2H cases confirms the greater stability of the 2H phase under the applied strain along different directions over the 1T$^{\prime}$ phase. 

\section{Conclusion}
In summary, we have performed first-principles density functional theory based calculations to investigate the modifications of geometrical, electronic and phononic properties of different polymorphs (2H and 1T$^{\prime}$) of single-layer MoS$_{2}$ under strain. A large number of strain profiles were considered for a comparison between the response of the semimetallic 1T$^{\prime}$ and the well-studied semiconducting 2H phase against the applied strain. Our calculations indicate that the changes in the electronic properties of both phases strongly depend on the direction of the applied strain. For the 2H phase, a direct to indirect band gap transition occurs at very low values of strain, and a semiconductor to metal transition occurs at a biaxial tensile strain of 10\%. In the 1T$^{\prime}$ phase, at different values of the applied strain, depending on the direction, the bulk states start to overlap with the edge states and thereby hamper the edge transport phenomena. For the 2H phase, the biaxial strain is seen to be the most effective for the semiconductor to metal transition, whereas, for the 1T$^{\prime}$ phase, the uniaxial strain along the direction of the zigzag chain of Mo atoms is found to be the most effective for the semimetal to metal transition. The strain-induced changes in the electronic structures are correlated with the corresponding change in the interaction strength between the Mo-d and S-p orbitals. We also investigated the strain-dependent lattice dynamics and lattice stability of monolayer MoS$_{2}$ in different crystalline forms. For both the 2H and 1T$^{\prime}$ phases, the tensile strain induces red-shift, and the compressive strain induces blue-shift of the Raman active modes. Also, for the 2H phase, the out-of-plane acoustic mode (ZA) becomes linear with the applied strain, which shows quadratic nature near the zone centre in the pristine condition. We analyzed the effect of strain on the stability of the crystal structures by looking at the emergence of the imaginary branches in the phonon dispersion curves. The 2H phase appears to have better stability against the applied strain compared to the 1T$^{\prime}$ phase, in which imaginary phonon modes emerge at a much lower strain of 3-4 \%. Our study, therefore, not only focuses on the importance of the strain application in tuning the electronic and the phononic properties of different polymorphs of monolayer MoS$_{2}$, but also provides an idea about the strength of the different phases and consequently the safe limit of the strain application. This work can further be extended to investigate the modulation of the thermoelectric behaviour with the application of strain. We hope our results will become very useful in further theoretical and experimental studies and the possible integration of single layer MoS$_{2}$ in tunable nanodevices.

\begin{acknowledgments}
	All first-principles calculations were performed using the Supercomputing facility of IIT Kharagpur established under National Supercomputing Mission (NSM), Government of India and supported by the Centre for Development of Advanced Computing (CDAC), Pune. SC acknowledge MHRD, India, for financial support.
\end{acknowledgments}

\bibliographystyle{apsrev4-2}
\bibliography{biblio}

\end{document}